\documentstyle[12pt]{article}
\title{The Half-Life and the Nuclear Matrix Element of the
$2\nu\beta\beta$ Decay in $^{76}Ge$ \thanks{The project supported
by National Natural Science Foundation of
China and Grant of Academia Sinica.}}
\author{Wu Hui-fang and Huang Tao  \\
Institute of High Energy Physics, Academia Sinica,\\
 P.O.Box 918, Beijing 100039, China}
\date{}
\begin{document}
\maketitle

\begin{abstract}
We apply the shell-model wave-functions of $^{76}Ge$ to calculate
the nuclear matrix elements and the half-life of the $2\nu\beta\beta$ decay
in $^{76}Ge$. Our result is comparable to the recent observed
$2\nu\beta\beta$-decay half-life of $^{76}Ge$. Furthemore it provides the
upper limit of the effective neutrino mass is about 0.4 eV from the recent
experimental data of the $0\nu\beta\beta$-decay half-life.
\end{abstract}

\vspace{0.5cm}


\vspace{1cm}

\section*{(I) Introduction}

{\hskip 0.6cm}The two-neutrino double beta $(2\nu\beta\beta)$ decay is allowed in the
standard theory of the electroweak interaction and has been observed in the
laboratory$^{[1]}$. A number of experiments have reported the direct observation
of the $2\nu\beta\beta$ modes. For instance, Elliott and collaborators
first observed the $2\nu \beta\beta$ decay of $^{82}Se^{[2]}$ in 1986. Ejili et al
and Alston-Garnjost et al have observed the $2\nu\beta\beta$ decay of
$^{100}Mo^{[3]}$. And the $2\nu\beta\beta$ decay of $^{76}Ge$ have been observed by
three groups-Russian$^{[4]}$, American$^{[5]}$ and Heidelberg-Moscow$^{[6]}$
groups. The future of double beta experiments will be dominated by the use of
enriched detectors. Among them, $^{76}Ge$ plays a particular favorable role.
Theoretically, one can predict the decay half-lifes with the calculated nuclear
matrix elements,since it is parameter-free from the particle physics side. The
experiment for the $2\nu\beta\beta$ decay will test the theoretical prediction
of the nuclear matrix elements.

\vspace{0.5cm}

However the situation is different for the neutrinoless double beta $(0\nu\beta\beta$
decay, where no neutrino is emitted and lepton number is violated. These
processes can only occur, as there is an exchange of Majorana neutrino.
Consequently the $0\nu\beta\beta$ decay half-life depends besides on the
nuclear matrix elements also on unknown parameters of the decay mechanism.

\vspace{0.5cm}

Furthermore, both modes are related to each other by the nuclear matrix
elements. Thus one can use the same wave functions of nuclei to calculate the
nuclear matrix elements for the $2\nu\beta\beta$ and $0\nu\beta\beta$-decays.
The availabe experimental data of the $2\nu\beta\beta$ decay will test $2\nu\beta\beta$
nuclear matrix elements and tell us how reliable the wave functions are. With
a more accurate wave-functions one can extract the neutrino mass information from the
experimental half-life limits of the $0\nu\beta\beta$ decay. The information about
the neutrino mass and neutrino mixing attracts the most attention due to the recent
experimental progress$^{[7]}$.

\vspace{0.5cm}

In this paper, following the shell-model wave-functions of $^{76}Ge$ and
$^{76}Se$ which were discussed in the previous publications$^{[8,9]}$, we
evaluate the half-life and the nuclear matrix elements of the $2\nu\beta\beta$
decay in $^{76}Ge$. The latest Heidelberg-Moscow measurement for the $2\nu\beta\beta$
decay of $^{76}Ge$ provides a test to these wave-functions. This comparison
confirms our nuclear wave-functions that apply to the $0\nu\beta\beta$ decay
of $^{76}Ge$. The result shows that the theoretical calculation from our
wave-functions is comparable with the experimental data of the $2\nu\beta\beta$
decay $^{76}Ge$ and the upper limit of the neutrino mass is less than 0.4eV from
the $0\nu\beta\beta$ decay of $^{76}Ge$.

\section*{(II) Calculated matrix element of $2\nu\beta\beta$ decay $^{76}Ge$}

{\hskip 0.6cm}The $2\nu\beta\beta$ mode has two electrons and two antineutrinos
in the final states and is expected to appear in the standard model in the
second order process of weak interaction, such as
\begin{eqnarray}
^{76}Ge \rightarrow ^{76}Se + 2e^{-} + 2\tilde{\nu}_{e} ~~~.
\end{eqnarray}
The decay amplitude associated with this process takes the form$^{[10]}$
\begin{eqnarray}
{\cal J}_{2\nu} &=& E_{m} \left [ \frac{<e_{1}e_{2}\nu_{1}\nu_{2}\psi_{f}
\mid H_{\beta} \mid e_{1}\nu_{1}\psi_{m} ><e_{1}\nu_{1} \psi_{m} \mid H_{\beta}
\mid \psi_{i}>}{E_{i} - E_{e_1} - E_{\nu_1} - E_{m}} \right.  \nonumber \\
&=& \left. \frac{<e_{1}e_{2}\nu_{1}\nu_{2}\psi_{f}
\mid H_{\beta} \mid e_{2}\nu_{2}\psi_{m} ><e_{2}\nu_{2} \psi_{m} \mid H_{\beta}
\mid \psi_{i}>}{E_{i} - E_{e_2} - E_{\nu_2} - E_{m}} - (e_{1} \leftrightarrow
e_{2}) \right ] ~~~~~, 
\end{eqnarray}
where $H_{\beta}$ is the hamiltonian of the weak interaction.
\begin{eqnarray}
H_{\beta} = \frac{G}{\sqrt{2}} \int J_{\mu}(x) L_{\mu}(x) d^{3}x ~~~~.
\end{eqnarray}
The $\psi_{i}, \psi_{m}$ and $\psi_{f}$ in Eq.(2) are the initial, intermediate and
final state wave-functions, respectively, $E_{i}$ and $E_{m}$ are the inital and
intermediate state energies. $G, L_{\mu}$ and $J_{\mu}$ in Eq.(3) are Fermi
coupling constant, the leptonic and hadronic currents respectively. $L_{\mu}$
and $J_{\mu}$ can be expressed as
\begin{eqnarray}
L_{\nu} = \bar{e}\gamma_{\mu}(1 - \gamma_{5}) \nu 
\end{eqnarray}
and
\begin{eqnarray}
J_{\mu} = \bar{N} \gamma_{\mu} (F_{V} - F_{A}\gamma_{5}) \tau_{+} N ~~~~~,
\end{eqnarray}
with $F_{V} = 1.0$ and $F_{A} = 1.25$.

\vspace{0.5cm}

Employing the usual closure approximation to the intermediate states,
the half-life $T^{2\nu}_{1/2}$ is given by
\begin{eqnarray}
T^{2\nu}_{1/2} = \frac{1n~2}{f_{GT} \mid M_{GT} \mid^{2}}
\end{eqnarray}
where
\begin{eqnarray}
M_{GT} = < \psi_{f} \mid \Sigma_{i<j} \tau_{+}(i) \tau_{+}(j) \sigma(i)
\sigma(j) \mid \psi_{i} > 
\end{eqnarray}
and
\begin{eqnarray}
f_{GT} &=& \xi_{2\nu} \frac{2m^{11}_{e} G^{4}}{7!\pi^{7}}
\left [\frac{F^{PR}(Z)}{E_{i} - \bar{E}_{m} - T_{0}/2 - m_{e}} \right ]^{2} \times
 \nonumber \\
&\times & \left [ 1 + \frac{\tilde{T}_{0}}{2} + \frac{\tilde{T}^{2}_{0}}{9} +
\frac{\tilde{T}^{3}_{0}}{90} + \frac{\tilde{T}^{4}_{0}}{1980} \right ] \times
\tilde{T}^{7}_{0} 
\end{eqnarray}
Here $\tilde{T}_{0} = T_{0}/m_{e}$ with $T_{0}$ being the total kinetic energy
of the outgoing electrons and $m_{e}$ being the electron mass. For $^{76}Ge,
T_{0} = 2.045 ~MeV. ~\bar{E}_{m}$ is the average energy of the intermediate
nuclear states. According to the statistics study of the $\beta$ decay,
the average intermediate energy is chosen reasonably and $\bar{E}_{m} - E_{i}
= 7.88 MeV$ in the case of $^{76}Ge^{[10]}. ~~F_{PR}(Z)$ is the nonrelativistic
Coulomb correction factor$^{[10]}$ and its use permits analytic evaluation of
the phase space integrals appearing in Eq.(8). $\xi_{2\nu}$ in Eq.(8) represents
the difference between this approximation and an exact integrals of the phase
space. For the $2\nu\beta\beta$ decay of $^{76}Ge, \xi_{2\nu} = 1.65^{[10]}$.

\vspace{0.5cm}

We have calculated the wave-functions of $^{76}Ge$ and $^{76}Se$ by using a simple
shell-model structure$^{[8,9]}. ~~Ni^{56}$ is adopted as an inert core for
$^{76}Ge$ and $^{76}Se$. The shell-model space for these nuclei includes four
single particle particle orbits $\{ 1 P_{3/2}, 0f_{5/2}, 1p_{1/2}, 0g_{9/2} \}$.
We employed the modified surface delta interaction (MSDI)$^{[11]}$ as the two
body residual interaction. The single-particle energies and first set of
parameters of parameters of MSDI in the Table 1 were taken from Faessler
et al$^{[12]}$. The second set of parameters of MSDI was used to the $2\nu\beta\beta$
decay of $^{82}Se$ in our previous paper$^{[13]}$.

\vspace{0.5cm}

\begin{center}
{\bf Table 1.} The nuclear matrix elements and the half-lifes of the $2\nu\beta\beta$
decay in $^{76}Ge$ Corresponding the different parameters of MSDI.
\end{center}

\begin{tabular}{ccccccccccc} \hline
  & & & & & & & & & & \\
 \multicolumn{5}{c}{MSDI} &\multicolumn{4}{c}{$\varepsilon_{i}$(MeV)} & & \\
  & & & & & & & & &$M_{GT}$ &$T^{2\nu}_{1/2}(th)$ \\ \cline{2-9}
  & & & & & & & & & \\ 
 &$A_{0}$ &$A_{1}$ &B &C &$1p_{3/2}$ &$0f_{5/2}$ &$1p_{1/2}$ &$0g_{9/2}$
 & &$\times 10^{21}(years)$  \\
 & & & & & & & & &  \\ 
  & & & & & & & & &  \\ \hline
1 &0.43 &0.35 &0.33 &0.00 &0.00 &1.75 &2.20 &3.39 &0.353 &5.17  \\ 
   & & & & & & & & &  \\ \hline
2 &0.31 &0.30 &0.30 &0.00 &0.00 &1.75 &2.20 &3.39 &0.479 &2.82  \\
 & & & & & & & & &  \\ \hline
\end{tabular}

\section*{(III) Comparison between the calculated result and the
experimental data}

{\hskip 0.6cm}The recent experimental data for $2\nu\beta\beta$-and
$0\nu\beta\beta$-decay modes of $^{76}Ge$ are given,
\begin{eqnarray}
T^{2\nu}_{1/2}(exp) = (1.77^{+0.01+0.13}_{-0.01-0.11} ) \times 10^{21} ~years^{[6]}
\end{eqnarray}
and
\begin{eqnarray}
T^{0\nu}_{1/2}(exp) > 5.7 \times 10^{25} ~years^{[7]}
\end{eqnarray}

The calculated result (see the table 1.) is comparable to the observed
$2\nu\beta\beta$-decay half-life of $^{76}Ge$ and the half-life from the
second line is close to the observed value in Eq.(9). Thereore, one can infer
that our wave-functions of $^{76}Ge$ and $^{76}Se$ may have more satisfied
construction with the reasonable parameter choices. With the same wave-functions
one can obtain that the effective neutrino mass is less than 0.4eV by using the
latest experimental lower limit (Eq.(10)) for the $0\nu\beta\beta$-deday
half-life of $^{76}Ge$.

\vspace{0.5cm}

As we know, the result of neutrino oscillation experiments on the effective
neutrino mass$^{[14]}$ gives the upper limit may be around $10^{-2}eV$. In
order to approach this limit from the nuclear $0\nu\beta\beta$ decay we
suggest to improve the existing experimental limit of the $0\nu\beta\beta$-decay
half-life by using the enriched $^{76}Ge$. More efforts to improve the theoretical
calculation on the nuclear matrix element of $^{76}Ge$ are also necessary.

\end{document}